\documentclass[11pt]{article}
\usepackage{geometry} % Allows the configuration of document margins
\usepackage{fancyhdr}
\usepackage[linguistics]{forest}
\usepackage{amsfonts}
\usepackage{amssymb}
\usepackage{setspace}
\usepackage{hyperref}
\usepackage{enumitem}
\usepackage{natbib}
\usepackage{comment}
\usepackage{txfonts}
\usepackage{pxfonts}
\usepackage{hyperref}
\usepackage{xcolor}

\title{On the Boundary of the Cosmos}

\author{Daniel Linford}

\begin{document}

\maketitle
\thispagestyle{fancy}

\begin{abstract}\noindent Intuitively, the totality of physical reality -- the Cosmos -- has a beginning only if (i) all parts of the Cosmos agree on the direction of time (the Direction Condition) and (ii) there is a boundary to the past of all non-initial spacetime points such that there are no spacetime points to the past of the boundary (the Boundary Condition). Following a distinction previously introduced by J. Brian Pitts, the Boundary Condition can be conceived of in two distinct ways: either topologically, i.e., in terms of a closed boundary, or metrically, i.e., in terms of the Cosmos having a finite past. This article proposes that the Boundary Condition should be posed disjunctively, modifies and improves upon the metrical conception of the Cosmos's beginning in light of a series of surprising yet simple thought experiments, and suggests that the Direction and Boundary Conditions should be thought of as more fundamental to the concept of the Cosmos's beginning than classical Big Bang cosmology.\end{abstract}

%\keywords{cosmology, big bang theory, philosophy of time, philosophy of spacetime}

\doublespacing

\section{Introduction}\label{sec1}

Despite the fact that the philosophical foundations of physical cosmology have seen a rapid growth spurt over the past decade, comparatively little work has been devoted to answering what, precisely, it would mean for the whole of physical reality -- herein, the Cosmos -- to have had a beginning of its existence. J. Brian Pitts \cite{Pitts:2008} has previously considered two proposals for a necessary (but not sufficient) condition for the Cosmos to have begun to exist. First, the \emph{Metrical Conception}, which roughly states that for the Cosmos to have begun to exist, the Cosmos must have a finite past. Second, the \emph{Topological Conception}, which roughly states that for the Cosmos  to have begun to exist, the Cosmos must have a past closed boundary, such that no part of the Cosmos existed before the boundary. Let's call the disjunction of the two conceptions -- that is, that for the Cosmos to have had a beginning, the Cosmos must satisfy the Metrical Conception or the Topological Conception -- the \emph{Boundary Condition}.

In this paper, I propose that a modified and improved version of the Boundary Condition be accepted as a necessary (but not sufficient) condition for the Cosmos to have had a beginning. I also offer a series of simple yet surprising ways in which a spacetime might (counterfactually) satisfy the two conditions. Moreover, I conjoin the Boundary Condition with an additional condition -- which I will call the Direction Condition -- and demonstrate a relationship between spacetime models satisfying the two conditions and classical models of the Big Bang.

In section \ref{Direc-Cond}, I summarize the Direction Condition, as previously offered in \cite{Matthews:1979} and \cite{Castagnino:2003}. In section \ref{top-conc-section}, I motivate the Topological Conception by considering metrical conventionalism and bimetric spacetimes. Following my development of the Topological Conception, in section \ref{Metrical-Conc}, I turn to considering the Metrical Conception. There, I develop three thought experiments (the Partially Amorphous Cosmos, the Fractal Cosmos, and the $\omega\omega^*$ Cosmos) in order to motivate the idea that the Cosmos could have a beginning, in the metrical sense, even if the Cosmos includes observers from whose perspective the Cosmos is either indeterminately or infinitely old. In subsection \ref{gap}, I turn to a discussion of the generalized affine parameter and developing a more sophisticated version of the Metrical Conception than has previously been offered. I consider some objections to my version of the Boundary Condition in section \ref{objections}. Lastly, in section \ref{big-bang}, I turn to a discussion of how the Direction and Boundary Conditions relate to classical Big Bang theory. As I demonstrate, if the Direction and Boundary Condition are conjoined with some other assumptions, one can \emph{derive} classical Big Bang theory. Although classical Big Bang theory will likely be replaced in a future quantum theory and so should not be taken as a realistic description of the Cosmos's origins, classical Big Bang theory is a prototypical example of a theory purporting to describe the Cosmos's origins. Since we can derive classical Big Bang theory from a set of assumptions that includes the Direction and Boundary Conditions, this should be taken to (i) provide evidence that I have correctly characterized two conditions for the Cosmos's beginning while also (ii) suggesting that the Direction and Boundary Conditions are more fundamental than classical Big Bang theory.

Before proceeding, I offer a few comments on some methodological desiderata that I assume throughout. While physicists and philosophers of physics often (though, of course, not always) assume a naive realist conception of relativity, the \emph{concept} of the Cosmos's beginning should not assume a naive realist conception of relativity. If a pre-relativistic conception of space and time had turned out to be correct, physicists and philosophers would still have had room to wonder whether the Cosmos had a beginning. Likewise, while a naive realist conception of relativity entails that spacetime is a continuum, the concept of the Cosmos's beginning should not assume that spacetime is a continuum and so should be consistent with the possibility that spacetime is discrete; moreover, while there are at least two conceptions of continua -- the Aristotelian and the Cantorian -- the concept of the Cosmos's beginning should not presuppose either conception of continua.\footnote{I will often speak as though spatio-temporal regions are point sets, but friends of the Aristotelian conception of continua can interpret at least some of my point set talk as involving a kind of pragmatic fiction. Moreover, the Aristotelian account of continua lacks the resources to distinguish between closed and open sets and, for that reason, do not have the resources to develop the Topological Conception, since the Topological Conception requires the notion of a point set with a closed boundary. However, the Boundary Condition is defined disjunctively and the second disjunct -- the Metrical Conception -- is the only natural way for Aristotelians to develop a conception of the Cosmos's beginning. For that reason, the Boundary Condition does not assume the truth of either the Aristotelian or Cantorian accounts.} Lastly, some philosophers \cite[pp. 183-184]{CraigSinclair:2009}; \cite[pp. 337-338]{Craig_Rel:1990}; \cite{Craig:2007}; \cite{Godfrey-Smith:1977}; \cite[p. 94]{Monton:2009}; \cite[p. 146]{Oderberg:2003}; \cite[pp. 135-136, 143, 147]{Mullins:2016}; \cite[p. 43]{Mullins:2011}; \cite[p. 62]{Leon:2019} have argued that \emph{beginning to exist} is a tensed notion, so that the Cosmos's beginning requires a tensed metaphysics of time. Likewise, e.g., Hans Reichenbach \cite[p. 11]{Reichenbach:1971} maintained that B-theory entails that nothing objectively begins or changes. Supposing B-theory does entail that nothing objectively begins or changes, then the Cosmos could have a beginning only if B-theory is false. For the purposes of this article, I will leave aside whether the Cosmos's beginning requires the truth of a tensed theory of time. Again, I am offering a necessary, but not necessarily sufficient, condition, so that, for all I say here, there may be other conditions required for the Cosmos to have had a beginning.

\section{\label{Direc-Cond}The Direction Condition}

Supposing that the Cosmos began, the beginning of the Cosmos must be temporally prior to all non-initial spacetime points the Cosmos includes. But in order for the beginning of the Cosmos to be temporally prior to all non-initial spacetime points the Cosmos includes, any hypothetical observer, located at any non-initial spacetime point, must agree that the putative beginning lies to their past. In order for any hypothetical observer, located at any non-initial spacetime point, to agree that the putative beginning lies to their past, all such observers would have to agree on the direction of time. In other words, the Cosmos could not include a beginning \emph{unless} the Cosmos includes a global direction of time. That global direction of time might be primitive, or defined by the direction of temporal becoming (supposing that $A$-theory is true), or by the global entropy gradient, or perhaps by something else. 

Regardless of how the global direction of time is determined, Matthews \cite{Matthews:1979} and Castagnino \cite{Castagnino:2003} have offered three conditions that the chronogeometric structure of a relativistic spacetime $(M, g)$ must satisfy in order for a global direction of time to be possible. Here, $M$ is a four dimensional manifold, that is, a set of points equipped with, e.g., topological structure, and $g$ is a spacetime metric with Lorentzian signature.\footnote{Although I will describe spacetime in terms of the pair $(M, g)$, I should not be interpreted as siding with what has sometimes been called the ``angle bracket school'' \cite{BrownRead:2022} or indeed any other approach to the foundations of spacetime theories; instead, I only intend that spacetime can be mathematically represented by the pair $(M, g)$.} Quoting from \cite[pp. 889--890]{Castagnino:2003}, the three conditions are:

\hfill

\begin{enumerate}
    \item $(M, g)$ is temporally orientable;
    \item For some $x \in M$, $(M, g)$ has a direction of time at $x$, that is, there is a non-arbitrary way of choosing the future lobe $C_x^+$ of the null cone $C_x$ at $x$;
    \item For all $x, y \in M$ such that $(M, g)$ has a direction of time at both $x$ and $y$, if the timelike vector $u$ lies inside $C_x^+$ and the timelike vector $v$ lies inside $C_y^+$ , then $u$ and $v$ have the same direction, that is, the vector resulting from parallel transport of $v$ to $x$ lies inside $C_x^+$.
\end{enumerate}

\hfill

\noindent  I will refer to the conjunction of (1)-(3) as the \emph{Direction Condition}. Since the Direction Condition includes three conjuncts, there are three ways that the Direction Condition could be violated. The first conjunct could be violated, that is, spacetime could fail to have a global direction of time by failing to be temporally orientable. The second conjunct could be violated, that is, given that spacetime is temporally orientable, there might nonetheless fail to be any objective way to assign a direction of time at every point in the spacetime manifold. Alternatively, the third conjunct could be violated, that is, spacetime could fail to have a global direction of time if an objective direction of time could not be assigned to every spacetime point without the direction varying from one spacetime region to another.

\section{\label{top-conc-section}The Topological Conception}

According to the perspective that I will adopt in this article, regions or durations -- whether of space, time, or spacetime -- can be correctly modeled as sets of points equipped with topological structure and perhaps also metrical structure. I do not mean to make a metaphysically thick assumption, that is, I am not, for example, supposing that spacetime is a substance composed by simples, i.e., points. Instead, I am supposing that -- regardless of whether substantivalism or relationalism or some other position provides the correct ontology of spacetime -- spacetime can be modeled as a point set.

\begin{comment}
    A point set is \emph{closed} just in case the point set includes its boundary points. For example, a segment of the real line from $0$ to $1$ is closed iff the segment includes both $0$ and $1$. A point set is \emph{open} just in case the point set does not include its boundary points. For example, the segment from $0$ to $1$ is open iff the segment does not include $0$ or $1$. And a point set is \emph{clopen} just in case the point set includes one boundary but not the other. For example, the segment from $0$ to $1$ is clopen iff the segment includes $0$ but not $1$ or includes $1$ but not $0$. The complement of any open set is closed. The union of a collection of open sets is open. Note that the notions of closed and open boundaries can be rigorously developed without appealing to any metrical notions, so that we can define the notion of a closed boundary without referring to the length of any curve.
\end{comment}

In order to motivate the Topological Conception of a beginning of the Cosmos, let's turn to a consideration of a view in the metaphysical foundations of spacetime theories called \emph{metrical conventionalism}.\footnote{For defenses of metrical conventionalism, see \cite{Poincare:2001b, Reichenbach:1958, Reichenbach:1971, Grunbaum:1968}.} According to metrical conventionalism, there are no non-conventional facts concerning the spacetime metric. The standard interpretation of relativity relativizes durations of time to reference frames. In this sense, relativity tells us that there is no fact about the duration of the temporal interval between two numerically distinct events independent of a choice of reference frame. The spacetime conventionalist goes one step further; for the conventionalist, the length of a given temporal interval cannot be specified even after we've specified a particular reference frame. For the conventionalist, after we've picked out a reference frame, we can determine the temporal duration between numerically distinct events (or spacetime points) only after selecting a specific convention for measuring temporal durations. If metrical conventionalism is true, there is no fact of the matter, independent of the adoption of a specific convention, as to the temporal duration that has passed so far in the Cosmos's history, including any fact about whether the temporal duration of the Cosmos's past history has been finite or infinite. Since, at the level of metaphysics, there are no conventional facts, metrical conventionalists say that there is no fact at all as to whether the Cosmos has a finite or an infinite past. Some authors have understood the question as to whether the Cosmos has a beginning to be synonymous with whether the Cosmos has a finite past. If the two questions are synonymous, then, since metrical conventionalism entails that the Cosmos's past is neither finite nor infinite -- but instead indeterminate -- metrical conventionalism entails that the Cosmos did not begin to exist.

For example, consider the half-open interval $(0, 1]$. Using the standard Lebesgue measure defined over the real line, the interval $(0, 1]$ has a length of $1$. But notice that $(0, 1]$ has the same set theoretic and topological features as $(-\infty, 1]$, that is, both intervals are continuous, half-open intervals containing an uncountable infinity of points; importantly, there exists a continuous bijection between the two intervals. If we set aside the Lebesgue measure -- that is, if we set aside the metrical features of the interval -- then there is no fact that distinguishes $(0, 1]$ from $(-\infty, 1]$ and so no fact distinguishing infinite from finite intervals. Likewise, suppose that the Cosmos has an open boundary to the past. In that case, the Cosmos's past history has the same topological features as a past eternal Cosmos.\footnote{My comments depend upon two facts: first, that there is no \emph{topological} feature that distinguishes the intervals $(0, 1]$ and $(-\infty, 1]$. Second, that there is a topological feature that distinguishes $(0, 1]$ and $[0, 1]$. What does it mean to say that there is no topological feature that distinguishes two intervals? $X$ and $Y$ are said to be topologically equivalent just in case there exists a continuous transformation from $X$ to $Y$ and which has a continuous inverse. In order to prove that $(0, 1]$ and $(-\infty, 1]$ are topologically equivalent, it suffices to construct a suitable continuous transformation and to show that the inverse of the transformation is continuous. I will assume that the topology on $(-\infty, 1]$ and $(0,1]$ is the subspace topology, that is, the topology inherited from the standard topology on $\mathbb{R}$. Consider the function $f(x) = -1/x + 2$. Trivially, $f(x)$ is a monotonically increasing function that maps the interval $(0, 1]$ to $(-\infty, 1]$; moreover, $f(x)$ trivially has a continuous inverse, at least on the interval in question. Having established the first result, I move to considering the second, namely, that there is a topological feature that distinguishes the intervals $(0, 1]$ and $[0, 1]$. Here, I will draw upon a more general result, namely, that there is a topological distinction between an open and a closed boundary. One can prove that compact (closed and bounded) sets can be mapped by continuous functions only to compact sets. Consequently, there is no continuous function mapping the compact set $[0,1]$ to the non-compact set $(0, 1]$; hence, $(0, 1]$ and $[0, 1]$ are not topologically equivalent \cite[p. 121]{Wapner:2005}. An anonymous referee claimed that the intervals $(0, 1]$ and $(-\infty, 1]$ are topologically distinct even though the two intervals are homeomorphic. To be sure, if we consider the two intervals as sub-intervals of $\mathbb{R}$, then there are topological features that the two intervals do not share. Here are three examples: (i) the interval $(-\infty, 1]$ is closed in the real numbers $\mathbb{R}$, whereas $(0, 1]$ is not; (ii) the complement of $(-\infty, 1]$ in $\mathbb{R}$ is connected, whereas the complement of $(0, 1]$ is not; (iii) the closure of $(0, 1]$ in $\mathbb{R}$ is compact, while the closure of $(-\infty, 1]$ is not. While one may use any one of these features to claim that $(-\infty, 1]$ and $(0,1]$ are not ``topologically equivalent'', all three are extrinsic features that depend upon how the intervals $(-\infty, 1]$ and $(0,1]$ are embedded within $\mathbb{R}$. Due to the fact that $(-\infty, 1]$ and $(0,1]$ are homeomorphic, there is no \emph{intrinsic} way to distinguish the two. Note that I am using the intervals $(-\infty, 1]$ and $(0,1]$ as ``stand-ins'' for the Cosmos's history; since the Cosmos is the totality of physical reality, the only features that matter for my purposes are the features that are intrinsic to an interval, and so, for my purposes, there is no relevant topological distinction between $(-\infty, 1]$ and $(0,1]$.} If metrical conventionalism is true and the Cosmos has an open boundary to the past, then the Cosmos did not have a beginning.

Nonetheless, intuitively, if the Cosmos includes a moment of time such that nothing at all existed prior to that moment of time, then -- so long as whatever other conditions are necessary for a beginning are satisfied (e.g., the Direction Condition) -- we should say that the Cosmos had a beginning. And this is so regardless of whether there are determinate metrical facts about the Cosmos.  For example, consider the closed interval $[0, 1]$. The interval $[0, 1]$ differs topologically from $(-\infty, 1]$ in virtue of having a closed boundary at $0$. Importantly, if we set aside all of the metrical features of the interval, we can still say that $[0, 1]$ has a closed boundary at the point we've labeled `$0$'. For analogous reasons, if spacetime conventionalism is true and the Cosmos has a closed boundary to the past of every non-initial spacetime point, we can still say that the Cosmos has a past boundary, even though there is no fact concerning the temporal interval between the boundary and ourselves. Intuitively -- supposing that the Cosmos satisfies whatever other conditions might be necessary for a beginning -- we come to the conclusion that the Cosmos would have a beginning in virtue of including a closed temporal boundary to the past of every non-initial spacetime point. This intuition might be thought to survive even if, as a naive realist interpretation of relativity entails, there are no global moments of time; for example, if our spacetime is globally hyperbolic and includes an initial Cauchy surface, that initial Cauchy surface might be understood as the Cosmos's beginning.

There is another closely related reason to prefer the Topological Conception over a conception that appeals to metrical information. Relativistic spacetimes are defined by a manifold $M$ and a metric $g$. $M$ is a collection of spacetime points equipped with topological structure. The spatio-temporal distance between any two points in $M$ can be defined in terms of $g$. There is no logical or mathematical inconsistency involved in defining a second distinct metric $g'$ over the same members of $M$, in terms of which we can define a second set of spatio-temporal distance relations. Theories that postulate two metrics on a given manifold are called \emph{bimetric} theories.\footnote{Bimetric theories indistinguishable from standard General Relativity have been considered in \cite{PittsSchieve:2003, PittsSchieve:2004, PittsSchieve:2007, Pitts:2019, Feynman:2003}; \cite[pp. 335-336]{Lockwood:2007}. A similar -- though in principle observationally distinguishable -- theory was considered in \cite{PittsSchieve:2007, Pitts:2018, Pitts:2019}; that theory approximates standard General Relativity arbitrarily well given a sufficiently small graviton mass. Bimetric theories have also been considered in \cite{Moffat:2003, Hossenfelder:2008, Hossenfelder:2016}.} And, of course, nothing at the level of logical or mathematical consistency forbids us from defining more than two metrics on the members of $M$; theories that postulate $n$ metrics on a given manifold can be called $n$-metric theories.

For an intuitive grasp of the notion of a bimetric theory, consider once more the half-open interval $(0, 1]$. Consider two points in that interval, for example, the points labeled by $0.5$ and $0.7$. On one way of defining the distance between the two points, the distance is the absolute value of the difference between their respective labels, i.e., $\vert 0.7-0.5 \vert = 0.2$. We can define another metric according to which the distance between any two points is the absolute value of the difference in the squares of the two labels, i.e., $\vert 0.7^2 - 0.5^2 \vert = 0.24$. We ordinarily think that the distance between two points has a unique value. But on a bimetric theory, there are two distances between any two points. In our example, the distance between the points labeled by $0.5$ and $0.7$ is \emph{both} $0.2$ \emph{and} $0.24$.

There are a variety of motivations for bimetric theories. As Henri Poincar{\'e} \cite[pp. 55-57]{Poincare:2001b} and Hans Reichenbach (e.g., \cite[pp. 30-34, 118-119]{Reichenbach:1958}) famously pointed out, any determination of chronogeometry will be systematically underdetermined. We can always save the hypothesis that spacetime has some specific chronogeometry by introducing forces that universally act on measuring instruments and distort all measurements taken by rulers or clocks. Poincar{\'e} and Reichenbach argued that, given our inability to determine which effects are legitimately chronogeometrical, there is no fact of the matter as to which effects are due to forces and which are due to chronogeometry. Philosophers of science have since given up on verificationism and are less prone to infer from systematic underdetermination between two hypotheses $h_1$ and $h_2$ that there is no fact of the matter as to which of $h_1$ or $h_2$ are correct. For that reason, we can rethink Poincar{\'e}'s and Reichenbach's point; perhaps we can distinguish between the effective metric handed to us by our observations and whatever metric legitimately describes our spacetime despite our observations. In that case, the true duration of past time could be systematically hidden from us precisely because the true metric would be epistemically inaccessible. In that case, we would have no right to infer from the Cosmos appearing to have a finite age that the Cosmos really does have a finite age. (Note that I am merely discussing this case as an epistemic possibility for the course of future inquiry and not endorsing it. There may be other extra-empirical theoretical virtues that would help us to distinguish hypotheses about physical chronogeometry, e.g., parsimony and the like.)

However, there is the possibility that if a bimetric (or $n$-metric) theory is true, both metrics might be epistemically available to us; moreover, consider the dynamical approach to the foundations of spacetime theories, as championed by Harvey Brown \cite{Brown:1993, Brown:1997, Brown:2005}, in co-authored work with Oliver Pooley \cite{BrownPooley:2006, PooleyBrown:2001}, and by Robert Disalle \cite{Disalle:2006}, or the closely related functionalist approach to spacetime theories, as championed by Eleanor Knox, e.g., \cite{Knox:2019}. According to the dynamical approach, in the context of theories with a fixed spacetime background, the chronogeometric structure of spacetime is determined by the laws and not vice versa. According to spacetime functionalism, the chronogeometric structure of spacetime is determined by whatever it is that, in a given spacetime theory, plays the role of explaining various inertial effects. On either view, chronogeometry is explanatorily downstream from the laws and not vice versa.

Suppose then that we adopt the perspective that chronogeometry is explanatorily downstream from the laws and that one empirically available metric is useful for describing some class of phenomena and another metric for another, distinct class of phenomena. In the previously discussed example, the distance between the points labeled by $0.5$ and $0.7$ is $0.2$ with respect to one metric and $0.24$ with respect to another. We can suppose that one metric is required by one set of physical phenomena, the other metric by the other set of physical phenomena, and that each metric plays the same functional role -- perhaps a role in explaining inertial effects -- for the phenomena to which each metric respectively applies. In that case, perhaps we should say that the points labeled by $0.5$ and $0.7$ are $0.2$ distance units apart in one respect and $0.24$ units apart in another respect and that neither distance is the one true distance. 

Just as two points can be two distinct distances apart if fundamental physical theory includes two metrics, so, too, the Cosmos may be finitely old with respect to one metric and infinitely old with respect to another metric. In that case, even supposing that both metrics could be empirically determined, if a beginning of the Cosmos requires a finite past, there may not be a determinate fact as to whether the Cosmos began (see, for example, \cite[p. 128]{SwinburneBird:1966}; \cite{sep-cosmology-theology, Milne:1948, Misner:1969, Roser:2016, RoserValentini:2017}).

If we set aside the metric and focus only on $M$, then we have set aside all facts about the duration of past time. $M$ is a point set that has set theoretic properties, such as cardinality, and topological properties, but not metrical properties. Since $M$ does not come equipped with metrical properties in itself, we cannot, by focusing only on $M$, mathematically distinguish between whether $(M, g)$ is a spacetime with an open boundary in the finite past or whether $(M, g)$ is a spacetime with an open boundary in the infinite past. However, $M$ is equipped, by construction, with topological structure. The distinction between an open and a closed boundary is a topological feature. Therefore, without appealing to any metrical facts, we can mathematically distinguish a spacetime with a closed boundary -- that is, a spacetime with a topological beginning -- from a spacetime without a closed boundary -- that is, a spacetime without a topological beginning.

To complete my discussion of the Topological Conception, I turn to unpacking three distinct families of ways for the Cosmos to have a topological beginning. As we will see, two of the families are counterintuitive and surprising. The first family has a topological beginning in the most intuitive sense; that is, all members of the first family are such that there is a single closed boundary to the past of all non-initial spacetime points. Consider, for example, flat (Minkowski) spacetime. Let's define a system of coordinates with respect to a reference frame $F$ and let's excise the portion of the spacetime below the line $t=0$. The resulting spacetime has a closed boundary at $t=0$ and so includes a shared closed boundary to the past of all non-initial spacetime points. If the spacetime also satisfies whatever other conditions there may be for having a beginning, then, intuitively, the spacetime's initial closed bounding surface is the spacetime's beginning.

We can now turn to the second family. The second and third family will be developed by drawing inspiration from a now famous theorem due to philosopher David Malament \cite{Malament:1977b}. According to Malament's theorem, for temporally orientable spacetimes that possess a local past/future distinction, the spacetime's topological, differential, and conformal (but not metrical) structure can be completely determined by specifying a class of continuous time-like curves. Since all classical spacetimes with a topological beginning satisfy the Direction Condition and so have a global direction of time, Malament's theorem is applicable to all of the classical spacetimes of interest in this paper. This suggests that we can construct all of the classical spacetimes with a topological beginning, up to but not including their metrical structure, by specifying a class of time-like curves.

Consider a globally hyperbolic spacetime $S$ with Cauchy surface $\Sigma$ and that satisfies the Direction Condition. I will assume that $\Sigma$ is a \emph{spacetime-wide surface}, that is, for every spacetime point $p$ not included in $\Sigma$, $p$ is included in the causal future or the causal past of at least one point in $\Sigma$. Moreover, let's suppose that any maximally extended time-like curve passing through $\Sigma$ measures infinite proper time to the past of $\Sigma$. For every time-like curve $\gamma$ passing through $\Sigma$, define a monotonically increasing and linear function of that curve's proper time, which I will call the \emph{age}. Construct the age such that the age has the following properties. First, at $\Sigma$, the age has a finite, positive value for every time-like curve. Second, for every time-like curve $\gamma$ whose age at $\Sigma$ is $a$, there exists another time-like curve $\gamma'$ that no where intersects $\gamma$ and whose age at $\Sigma$ is $a+\varepsilon$, where $\varepsilon \in \mathbb{R}$ and $0 < \varepsilon < \infty$. Third, every time-like geodesic that intersects both the boundary and $\Sigma$ has a unique age. Now I will construct another spacetime $S'$, from $S$, by the following procedure. For any given time-like curve $\gamma$, remove from $S$ the points on $\gamma$ where the age is less than zero. Note that since we've removed the points where the age is less than zero, but not the points where the age is zero, each time-like curve has an initial point and so a closed boundary. In this case, even though every time-like curve in the spacetime had a beginning at some time in the finite past and a closed boundary to the past of $\Sigma$, there is generally no closed space-like boundary shared by all time-like curves in the entire spacetime.\footnote{This claim is easy to motivate. Consider, for example, the case where we begin with Minkowski space and then construct a new spacetime $S'$ using the procedure I've described. And now suppose, for reductio, that there exists a closed space-like boundary $\mathcal{B}$ shared by all of the time-like curves in $S'$. There are two cases that we can consider: first, the case where $\mathcal{B}$ is ``parallel'' to $\Sigma$, that is, the case where all of the time-like geodesics connecting $\mathcal{B}$ and $\Sigma$ measure the same proper time at the point of intersection with $\Sigma$, and, second, the case where $\mathcal{B}$ is not parallel to $\Sigma$. In the first case, since all of the time-like geodesics measure the same proper time when they intersect $\Sigma$, we are guaranteed that there are at least two time-like geodesics that measure the same age at the point at which they respectively intersect with $\Sigma$. But, by construction, no two time-like geodesics have the same age at their respective points of intersection with $\Sigma$. In the second case,  $\mathcal{B}$ is tilted with respect to $\Sigma$; in that case, there is some volume where $\mathcal{B}$ and $\Sigma$ intersect. Past the volume where $\mathcal{B}$ and $\Sigma$ intersect, $\Sigma$ does not exist. For that reason, the spacetime will include at least one point that is not in the causal future or the causal past of some point in $\Sigma$, so that $\Sigma$ is not actually a spacetime-wide surface. Again, one of our assumptions has been violated. Thus, since the spacetime's closed boundary cannot be a space-like surface in either case and the two cases are mutually exclusive and exhaustive, the spacetime's closed boundary cannot be a space-like surface. While this result is much more difficult to establish when we allow for arbitrary spacetime curvature, a single example suffices to establish that one can construct spacetimes with the features I've described.} Importantly, we can always trace the history of the spacetime further back -- in the sense that there is always a time-like curve that, relative to $\Sigma$, extends further to the past -- so that there is no specific time, i.e., no specific time-like surface, at which the Cosmos began. That is, there are examples of spacetimes where every object in the spacetime began to exist, but there is no \emph{one} time (or one space-like surface) that could even count as a candidate for the beginning of the spacetime as a whole.

In the thought experiment in the previous paragraph, we considered the set of time-like curves piercing $\Sigma$. We can construct a spacetime using the class of time-like curves that pierce $\Sigma$ and, given Malament's theorem, that class of curves will suffice for determining the non-metrical structure of one family of classical spacetimes with a topological beginning. And this suggests a class of spacetimes with a ``jagged'' closed boundary, so that, in some sense, one's distance from the beginning of the Cosmos depends upon where one resides within the Cosmos. However old the Cosmos is in one's own ``neck of the woods'', there may be some other spacetime point in $\Sigma$ where the Cosmos is older.

Let's turn to a third family of classical spacetimes with a topological beginning. Once again, consider a classical, globally hyperbolic spacetime with a Cauchy surface $\Sigma$. Let's repeat the same procedure as for the previously constructed spacetime, except with some modifications to the age function. Recall that for each time-like curve $\gamma$, the age is a linear and monotonic function of the proper time $t$, i.e., $age_{\gamma}(t) = bt + c$. Moreover, let's assign each time-like curve the index $\varepsilon$, where $\varepsilon$ is a real number between $0$ and $\infty$; assign the values of $\varepsilon$ such that there is a unique time-like curve for every positive real number and vice versa. \footnote{That is, $\varepsilon$ is a bijection from the positive real numbers to the set of time-like curves. We are guaranteed that this bijection exists because there is a bijection from $n$-dimensional space to the set of positive real numbers. Nonetheless, bijections between $n$-dimensional space and the positive real numbers are not generally smooth, thereby lending another reason why the boundary described is ``jagged''.} Let's say that $a_{\varepsilon}(\Sigma)$ is the value of the age for the curve $\varepsilon$ at the point where $\varepsilon$ intersects $\Sigma$. Now define each curve's respective age such that each curve's age, at the point where the curve intersects $\Sigma$, is a function of $\varepsilon$:

\begin{equation}
    a_{\varepsilon}(\Sigma) = bt(\Sigma) + c = \frac{1}{1+e^{-\varepsilon}}
\end{equation}

\noindent where $t(\Sigma)$ is the value of the proper time on curve $\varepsilon$ when $\varepsilon$ intersects $\Sigma$. Lastly, repeat the same procedure as before for removing points from the spacetime, i.e., along each curve, remove the points from the spacetime where the age is less than zero.

Notice that in the limit that $\varepsilon$ increases without bound, $a_{\varepsilon}(\Sigma)$ approaches $1$. No time-like curve has an age greater than $1$ at $\Sigma$ and yet no time-like curve has the greatest proper time at the point where that curve intersects $\Sigma$, i.e., there may be no oldest particle in the entire spacetime. In this case, there is no closed boundary shared by all time-like curves, since there is no Cauchy surface on which all time-like curves originate, but the spacetime is still bounded to the past because there is a maximum value in the duration of past proper time for every time-like curve. One may have the intuition that this family of spacetimes has a beginning in a stronger sense than the first family of classical spacetimes with a topological beginning that we examined. Indeed, this is so, because the spacetime has a boundary in two senses. On one hand, as already explained, the spacetime has a closed boundary because every time-like curve includes an initial boundary point. On the other hand, the spacetime has a boundary in the sense that there is a maximum value for the age that a particle could have when that particle passes through $\Sigma$; that is, no particle can have an age at $\Sigma$ greater than $1$.

Unfortunately, although we can mathematically distinguish spacetimes with a topological beginning from spacetimes without a topological beginning, we cannot, in general, \emph{empirically} distinguish the two. The only features of spacetime that can be empirically discovered are those related to the distribution of the matter-energy populating spacetime. Yet, in General Relativity, there is no relationship between closed boundaries and the matter-energy distribution. Alas, all hope is not yet lost for beings in a classical spacetime to discover their spacetime's beginning. In the next section, I turn to another conception of a past boundary and we will have to ask whether that conception allows us to tie the matter-energy distribution to the existence or absence of a boundary.

\section{\label{Metrical-Conc}The Metrical Conception}

Suppose that time is absolute and has an open boundary to the finite past such that there is no time before the boundary at which the Cosmos exists. Since the boundary is open, if there were no other sense in which the Cosmos could have a beginning than the Topological Conception, we should say that the Cosmos did not begin to exist. Nonetheless, there is a strong intuition that one way for the Cosmos to begin to exist would involve time having an open boundary in the finite past. And there is a strong intuition that if the Cosmos had another kind of open boundary -- namely, an open boundary infinitely far to the past of all spacetime points -- then the Cosmos did not begin to exist. Since this intuition concerns the lapse (or total duration) of past time, following Pitts \cite{Pitts:2008}, we can call the resulting conception of the Cosmos's beginning the \emph{Metrical Conception}.

In this section, I construct a new Metrical Conception of the beginning of the Cosmos. As a first pass, to be later modified in light of three thought experiments, the Metrical Conception states that the Cosmos began to exist in the metrical sense only if, for any arbitrarily designated interval of time $T$, with total duration $m(T)$, there are only a finite number of past non-overlapping intervals with duration $m(T)$. Let's call this initial version the \emph{Naive} Metrical Conception, or NMC. Below, I offer three novel thought experiments which help to pump the intuition that, contrary to the NMC, even if the Cosmos has an infinite past, the Cosmos may still have had the kind of beginning that ought to be covered by the Metrical Conception. In addition to providing an intuitively plausible analysis of the three thought experiments, the new Metrical Conception will fulfill two desiderata. First, the Metrical Conception should be \emph{consistent} with -- without entailing or committing to -- the view that a finite past, together with the Direction Condition, suffice for establishing that the Cosmos had a beginning. I am not claiming that merely having a finite past and satisfying the Direction Condition do suffice for having a beginning; there may be other conditions that are necessary for the Cosmos to have had a beginning. Second, there should be cases where the new Metrical Conception agrees with our intuition that the Cosmos had no beginning.

\subsection{Three Thought Experiments}

I now turn to a consideration of the three thought experiments. Before explicating the three thought experiments, I briefly describe a collection of preliminary mathematical notions. We can represent the points in an $n$-dimensional space recursively by taking Cartesian products of the real numbers, e.g., $\mathbb{R}^n \equiv \mathbb{R}\times...\textnormal{(n-2 times)}...\times\mathbb{R}$. The real numbers merely label points; to define the distance between two points, we need to define one or more metrical relations on $M$ as well as the ``lengths'' of some appropriate set of curves connecting the two points. For example, in relativistic spacetimes, $g$ is a rank $2$ tensor, with components $g_{\mu\nu}$, from which we can compute the ``distance'' (that is, the interval) between points $p$ and $q$ by maximizing $\int_p^q \sqrt{g_{\mu\nu} dx^{\mu}dx^{\nu}}$, where the integral is computed along a path from $p$ to $q$. Relativistic spacetimes can be modeled by a pair of objects, i.e., the manifold $M$ and the metric tensor $g$. We can provide an analogous, albeit anachronistic, description for pre-relativistic spacetimes. Newtonian and Galilean spacetimes are modeled using a manifold $\mathbb{R}^4$, a temporal metric $t$, describing the duration between any two instants of time, and a spatial metric $h$, describing the spatial distance between any two points in space. In Newtonian spacetime, points of space persist over time -- which can be represented by re-identifying the same spacetime points at successive times -- whereas, in Galilean spacetime, points do not persist over time.

Before continuing on to a discussion of the three thought experiments, I need to introduce a general principle that I will use to reach the lessons that I take from each of the thought experiments. Given any two observers $A$ and $B$, if the Cosmos began for $A$ then the Cosmos began for $B$ and vice versa. If a version of the Boundary Condition entails that the Cosmos began relative to some observer and did not begin relative to some other observer, then that version of the Boundary Condition is inadequate. Having laid out some mathematical foundations and stated a general principle, I continue on to a discussion of the three thought experiments.

\subsubsection{The Partially Amorphous Cosmos}

Some cosmological models include a spacetime region where there are no metrical facts and another spacetime region where there are metrical facts. Consider Bradford Skow's \cite{Skow2010} argument that an objective spacetime metric might not be either an intrinsic feature of spacetime or wholly the result of features intrinsic to spacetime. Instead, Skow argues, spacetime might have an objective, but extrinsic, temporal metric just in case there is some $x$ that plays the functional role, in the physical laws, of determining the ratios between any two non-overlapping spatio-temporal intervals.\footnote{Skow cashes out his view in terms of absolute time, but indicates that he intends for his view to be generalizable to relativistic spacetimes.} If metrical facts require a specific functional role to be fulfilled, then, in spacetime regions where that functional role is not fulfilled, there might not be any metrical facts, even though metrical facts do obtain in other spacetime regions.

For example, in Roger Penrose's \cite{penrose:2012} Conformal Cyclic Cosmology, there are no facts about spatio-temporal scale, that is, no metrical facts, at early or late times in the history of the observable universe.\footnote{On some quantum gravity theories -- such as causal set theory \cite{BombelliLeeMeyerSorkin:1987, Dowker:2006, Dowker:2013, Dowker:2017, Dowker:2020, BrightwellGregory:1991} -- the spacetime metric appears only in the theory's continuum limit, thereby allowing for the possibility that there are regions of the Cosmos where the spacetime metric is inapplicable. However, we should not necessarily think of those regions as amorphous in the sense discussed in this section. Consider, for example, Brightwell and Gregory's \cite{BrightwellGregory:1991} construction of the continuum limit for a spacetime interval spanned by a number of spacetime atoms ``linked'' together in a chain. When the chain is sufficiently long, the spacetime interval is proportional to the number of links in the chain. As causal set theorists like to say, in causal set theory, metrical facts are determined by counting. For that reason, supposing that there are only a small number of spacetime atoms in some region, so that the continuum limit does not apply in the region, we need only consider a larger region to recover relevant metrical facts. In any case, recall that the Boundary Condition for the Cosmos to have a beginning is disjunctive. If the initial portion of the Cosmos is correctly described by causal set theory, then, since causal sets always have closed boundaries, the Cosmos would satisfy the first disjunct -- by having a topological beginning -- and so would have a beginning.} A temporal (or spatio-temporal) interval for which there is no fact concerning the length of the interval -- that is, an interval to which metrical facts are inapplicable -- is said to be \emph{amorphous}. To put the view another way, if spacetime is metrically amorphous within some region, then there is no objective fact about the ratio of the durations of two non-overlapping intervals within that region. In relativistic spacetimes, lengths and temporal intervals depend upon the adoption of a specific reference frame. Amorphous time goes one step further in that if time is amorphous then, even within a given reference frame, there are no facts about how long a given temporal interval is. One example of amorphous time is time for which metrical facts are purely conventional, as already discussed, though amorphous time can also be such that one cannot even adopt a conventional metric. For the sake of simplicity, let's suppose that Newton and Galileo were correct that time is absolute.\footnote{\label{relativistic-amorph}Nothing crucial in this example hangs on whether time is absolute. The example can be reconstructed for relativistic spacetimes. To construct a relativistic spacetime without metrical structure, first consider a spacetime $S$ with metric $g_{\mu\nu}$. And now consider the metric $\tilde{g}_{\mu\nu}$ produced from $g_{\mu\nu}$ by the conformal transformation $\tilde{g}_{\mu\nu} = \Omega^2 g_{\mu\nu}$ where $\Omega$ is a positive and smooth but otherwise arbitrary scalar function. For relativistic spacetimes, multiplication by $\Omega^2$ leaves the spacetime's light cone structure unaltered. Call the resulting spacetime $\tilde{S}$. Two spacetimes that are related by such a transformation, e.g., $S$ and $\tilde{S}$, are said to be \emph{conformally equivalent}. A spacetime \emph{without} metrical structure can then be constructed by identifying all of the members of a given class of conformally equivalent spacetimes. Let's call the spacetime that results from identifying all of the members of a given class of conformally equivalent spacetimes $S_C$. Since the conformal transformation left the light cone structure unaltered, $S_C$ is equipped with light cone structure but not metrical structure and so $S_C$ is an example of a relativistic metrically amorphous spacetime. To construct a relativistic spacetime analogous to the spacetime inhabited by Pam and Jim, one can ``glue'' a metrically amorphous spacetime region $R$ between two regions that are not metrically amorphous.} Let's also suppose that there is a finitely long interval of non-amorphous time labeled $A$, followed by an interval of amorphous time labeled $B$, and then followed again by an interval of non-amorphous time labeled $C$. Formally, we are supposing that $A$ is a Newtonian or Galilean spacetime region with an objective temporal metric, that $B$ has the topology of a Newtonian or Galilean spacetime region without an objective metric, and that $C$ is another Newtonian or Galilean spacetime region. Suppose, further, that the Cosmos does not exist prior to $A$. Call this construction the \emph{Partially Amorphous Cosmos}.

Suppose that Pam is an arbitrarily chosen observer in $A$. Pam should say that the Cosmos began in her finite past. Suppose that Jim is an observer in $C$. For Jim, since there is an interval of amorphous time between himself and the beginning identified by Pam, there is no fact concerning how far in the past the Cosmos began. Consequently, even though, intuitively, Jim should agree that time began, there is no fact about how many isochronous intervals can be placed into Jim's past. Since there is no fact about how many isochronous intervals can be placed into Jim's past, the NMC entails the intuitively wrong conclusion; according to the NMC, the Partially Amorphous Cosmos did not begin to exist. A Newtonian or Galilean Cosmos with a beginning can have a non-initial segment in which there is no objective temporal metric. Instead of articulating the Metrical Conception in terms of there being a determinate number of isochronous intervals to the past, the Metrical Conception should entail that, for spacetimes with a metrical beginning, time is not metrically amorphous in the initial segment of the Cosmos's history.\footnote{A similar point has been previously made in various places, but, in particular, see \cite[pp. 125-126, 131]{Earman:1977}. As Earman argues, in order for the Cosmos to have an objectively finite past, there must be an objective best choice of global time function according to which the past is finite. In contrast, Hermann Weyl \cite{Weyl:1997} maintained that the choice of time scale is, is to a certain degree, conventional. In more technical terms, Weyl argued that there is gauge freedom in one's choice of metric tensor so that the metric tensor is determined only up to a conformal factor, as in footnote \ref{relativistic-amorph}. In any case, were Weyl's theory correct, time scale would not correspond to any objective physical fact \cite[p. 451]{penrose:2004}. Additional technical details for Weyl's theory can be found in \cite{sep-weyl}.}

\subsubsection{The Fractal Cosmos} 

There are a number of fractal curves whose arc length is infinite \cite[p. 38]{Koch:2018}; \cite{Mandelbrot:2018}, even though they occupy a finite region of the plane. Consider, then, a fractal curve with infinite arc length that occupies a region of the $x-y$ plane with an end point at the left at $x=-1$ and an end point at the right at $x=1$. We can ``glue'' a finitely long line segment, parallel to the $x$-axis, to the curve's left end point and another to the curve's right end point. Call the line segment on the left $L$, the fractal curve $C$, and the line segment on the right $R$. Restricting ourselves to the resulting $L-C-R$ compound geometric object, notice that:

\begin{enumerate}
    \item\hspace*{1sp} There is a finite distance between any two points in $L$.
    \item\hspace*{1sp} There is a finite distance between any two points in $R$.
    \item\hspace*{1sp} There is an infinite distance between any point in $L$ and any point in $R$.
\end{enumerate}

\noindent Once more, for the sake of simplicity, suppose that Newton and Galileo had been right that time is absolute.\footnote{An analogous construction can be produced using relativistic physics.} Moreover, suppose that absolute time had the metrical structure of $L-C-R$. Suppose that Pam is an arbitrarily chosen observer in the $L$ segment of history. By construction, $L$ is a finitely long line segment; for that reason, there is only a finite period of absolute time to Pam's past. Intuitively, the Metrical Conception, when conjoined with the fact that there is only a finite period of absolute time to Pam's past, should strongly suggest that there was a beginning of absolute time. However, for any arbitrarily chosen observer -- call them Jim -- in the $C$ or $R$ segments of history, the beginning suggested by Pam is located infinitely far in the past. For that reason, the NMC clashes with intuition for two reasons. First, since an infinity of finite isochronous intervals can ``fit'' into Jim's past, the NMC implies that the Fractal Cosmos does not have a beginning in the metrical sense. This is clearly the wrong conclusion; the fact that there is an infinity of finite isochronous intervals in Jim's past should not preclude Jim from concluding that his Cosmos has a beginning. Second, whether the Metrical Conception is satisfied should not be observer relative; nonetheless, Pam and Jim reach inconsistent conclusions as to whether their Cosmos satisfies the NMC.

\subsubsection{The $\omega\omega^*$ Cosmos} 

In this section, I construct what I will call the $\omega\omega^*$ Cosmos. Let's begin by considering the series of positive integers in increasing order: $1$, $2$, $3$, ... This sequence has order type $\omega$. Sequences of order type $\omega$ do not have a last member, but we can add in a last member $z$ by defining $z$ such that $z$ comes after every member in the sequence. We can then use $z$ to define a new sequence: $1$, $2$, $3$, ..., $z$. We can also consider the series of negative integers in increasing order: ..., $-3$, $-2$, $-1$. This sequence has order type $\omega^*$. Sequences with order type $\omega^*$ have no first member -- since the sequence of negative numbers has no start -- but we can add in a first member $a$ by defining $a$ such that $a$ comes before every member in the sequence. We can then use $a$ to define a new sequence: $a$, ..., $-3$, $-2$, $-1$. Lastly, we can ``glue'' together the $\omega$ and $\omega*$ sequences by identifying $z$ with $a$: $1$, $2$, $3$, ..., $z$, ..., $-3$, $-2$, $-1$. Call this the $\omega\omega^*$ sequence. Given a countably infinite set of points, we can identify each point in the set with one member of the $\omega\omega^*$ sequence and we can define topological and ordinal relations such that points labeled by sequential values in the $\omega\omega^*$ sequence are neighbors, e.g., the point labeled $1$ is to the right of the point labeled $2$ and all of the points labeled by negative integers are to the right of the positive integers.

We can define a metric over the point set labeled by the $\omega\omega^*$ sequence with the following properties: (i) the distance between any two points in the portion labeled by the positive integers is given by the absolute value of the difference between the corresponding two integers, (ii) the distance between any two points in the portion labeled by the negative integers is given by the absolute value of the difference between the corresponding two integers, (iii) the distance between $z$ and any other point is infinite, and (iv) the distance between any of the points labeled by a positive integer and any point labeled by a negative integer is infinite. We've succeeded in defining a point set equipped with topological structure, that is, a manifold, and a metric. While the Cosmos may still fail to satisfy other conditions for having a beginning, the mere fact that past time is infinite should not bar us from saying that the Cosmos began to exist. Time would have a first instant, namely, the point labeled by $1$ that, possibly given other conditions, would be natural to call time's beginning.

The NMC reaches conclusions for the $\omega\omega^*$ Cosmos that clash with our intuitions for reasons parallel to those we identified for the Fractal and Partially Amorphous Cosmoi. On one hand, the NMC implies that the $\omega\omega^*$ Cosmos does not have a metrical beginning for one set of observers. For any observer in the $\omega^*$ portion, an infinitude of finite isochronous intervals can ``fit'' into their past; in turn, for that set of observers, the $\omega\omega^*$ Cosmos does not satisfy the NMC. Nonetheless, there is a strong intuition that the $\omega\omega^*$ Cosmos has a metrical beginning. On the other hand, the NMC delivers inconsistent answers for whether the $\omega\omega^*$ Cosmos satisfies the condition; given the NMC, for any observer in the $\omega$ portion, the Cosmos has a metrical beginning, whereas, for any observer in the $\omega^*$ portion, the Cosmos does not have a metrical beginning. According to the new Metrical Conception that I develop below, supposing the $\omega\omega^*$ Cosmos satisfies the other conditions necessary for having a beginning, the $\omega\omega^*$ Cosmos's beginning is not relative to any set of observers.

\subsubsection{Drawing lessons from the three thought experiments}

In the cases of the Fractal Cosmos and the $\omega\omega^*$ Cosmos, one may object that infinity is not a number, in which case distances and temporal intervals cannot be infinite. Four replies can be offered. First, that there are spacetime points between which the temporal interval is not well-defined would suffice for my purposes. For that reason, if we understand the temporal intervals involved not as infinite but as divergent -- and so as not well-defined -- similar conclusions follow. Second, while infinity is not a \emph{real} number, there are well known geometrical constructions in which points are included that are at an infinite distance from other points. One family of constructions are the fractal curves already discussed. For another example, consider the projection of the Riemann sphere on to the complex plane, which allows one to identify complex infinity with the sphere's north pole. There is no recognized \emph{mathematical} difficulty involved in including ``points at infinity'' in a given construction. Whether there are metaphysical problems involved in such constructions has -- in my view -- yet to be successfully shown. Third, there are solutions to the Einstein Field Equations -- such as Malament-Hogarth spacetimes -- that include observers who, in finite time, can observe the results of a computation that takes infinite time to perform \cite{Hogarth:1992, EarmanNorton:1993, sep-spacetime-supertasks, EtesiNemeti:2002}. (On a more technical level, what's crucial about Malament-Hogarth spacetimes is the feature that a time-like half-curve, along which there is infinite proper time, can ``fit'' inside some observer's past light cone, where the observer is not located at time-like infinity.) If we accept some standard solutions to the Einstein Field Equations, e.g., Kerr black holes or anti-De Sitter spacetime, as legitimately metaphysically possible, then we need to allow for the metaphysical possibility of infinite arc lengths. Fourth, supposing that one considers the Fractal Cosmos and the $\omega\omega^*$ Cosmos as metaphysically impossible, one is still left with the Partially Amorphous Cosmos as a viable epistemic possibility.

According to a closely related objection, one might think that the thought experiments I've presented are metaphysically impossible because the thought experiments include instants that (i) are not at the beginning or end of time but which (ii) have no immediate preceding or succeeding instant. For example, in the $\omega\omega^*$ Cosmos, there is a point ``at the middle'' of time, i.e., the point labeled $z$, with no immediate preceding or succeeding instant. In reply, note that on any view according to which time is a Cantorian continuum, \emph{all} instants of time have no immediate preceding or succeeding instant. While there may be objections to the view that time is a Cantorian continuum, the $\omega\omega^*$ Cosmos (for example) is at least no worse, in that respect, than the view that time is a Cantorian continuum. While this article takes no position on whether time is a Cantorian continuum, the notion that the Cosmos had a beginning should at least be consistent with time being a Cantorian continuum. Moreover, note that this objection applies only to my specific construction of the $\omega\omega^*$ Cosmos; we can construct another, qualitatively similar Cosmos in which every non-initial point, including $z$, has an immediate predecessor and an immediate successor, i.e., $1, 2, 3, ..., \omega-2, \omega-1, \omega, \omega+1, \omega+2, ..., -3, -2, -1$.

\begin{comment}
According to Craig and Sinclair's account, the Cosmos began only if the Cosmos's history includes no more than a finite number of isochronous intervals earlier than any arbitrarily chosen interval. In the three thought experiments, so long as the Cosmos satisfies the Direction Condition (and possibly other conditions, such as the truth of a tensed theory of time, for the Cosmos's beginning), there is a strong suggestion that the Cosmos began; nonetheless, in the three thought experiments, there are some temporal intervals such that there is no finite or determinate number of isochronous intervals to that interval's past. According to Swinburne's account, the Cosmos began just in case there are a finite number of isochronous intervals earlier than the present. But in the three thought experiments, there could be observers situated such that there there is no finite (or determinate) number of isochronous intervals before their present. Moreover, since Swinburne's version of the Metrical Conception is indexed to the present of a given observer, Swinburne's version yields inconsistent conclusions about whether a given Cosmos had a beginning. Craig, Sinclair, and Swinburne's accounts -- like my three thought experiments -- assume that time is absolute. 
\end{comment}

The three thought experiments that I've considered allow me to identify two problems with the NMC. First, the NMC is indexical and so yields inconsistent results as to whether each Cosmos satisfies the NMC. As a first pass for removing the indexical, we can say that the Cosmos began to exist in the metrical sense only if, for any arbitrarily chosen moment of time $\tau$, and for any arbitrarily designated interval of time $T$, with total duration $m(T)$, there are only a finite number of intervals before $\tau$ with duration $m(T)$. Unfortunately, this first pass won't work either; this first pass entails that the three Cosmoi are beginningless, since there are moments of time in all three whose past are either indefinitely or infinitely long.

As a third pass, the new metrical conception might say that the Cosmos has a beginning in the metrical sense only if (i) there is a (closed or open) boundary $B$ to the past of all spacetime points and (ii) there exists \emph{some} time $\tau$ such that, according to the objective metric of absolute time, the span of time between $B$ and $\tau$ is finite. Like my three thought experiments, this third pass assumes that time is absolute, whereas the metrical conception should be consistent with relativistic physics.\footnote{Ideally, the account should also be consistent with a future quantum gravity theory, but, given that we do not yet possess a successful quantum gravity theory, the account that I offer here will need to be provisional.} For that reason, a Metrical Conception of the beginning of the Cosmos that did not assume absolute time is desirable. However, this third pass highlights a feature that I will pursue below, namely, that the Cosmos satisfies the metrical conception just in case the Cosmos includes an initial finite segment of the Cosmos's history. Much of what follows in the next section concerns what it means for the Cosmos to include an initial finite segment. In order to construct the notion of an initial finite segment, and so a new version of the Metrical Conception that does not assume absolute time, I need to first explicate the generalized affine parameter.

\subsection{\label{gap}The Generalized Affine Parameter}

In place of an absolute time for the whole of the Cosmos, relativistic physics introduces proper time, which is parametrized along individual trajectories. Perhaps the notion that the Cosmos has a finite past can be re-phrased in terms of proper time. For example, consider two spacetime-wide space-like surfaces $\Sigma$ and $\Sigma'$. Intuitively, if every object with a trajectory $\gamma$ that passes through both $\Sigma$ and $\Sigma'$ records that there is only finite proper time between the point where $\gamma$ intersects $\Sigma$ and the point where $\gamma$ intersects $\Sigma'$, then we can say that there is a finite portion of the Cosmos's history between $\Sigma$ and $\Sigma'$. This procedure backfires when we consider that the proper time along any light-like trajectory is zero. Consequently, even if every massive particle records the Cosmos's past history as infinite, light will record no duration at all.

To resolve this difficulty, we need a suitable alternative $\lambda$ to proper time with two features. First, for bodies moving slower than light, $\lambda$ should distinguish infinitely from finitely long trajectories. That is, trajectories along which there is finite proper time should be assigned finite values of $\lambda$ and trajectories along which there is infinite proper time should be assigned infinite values of $\lambda$. Second, $\lambda$ should parametrize the points along the trajectories followed by light in such a way that numerically distinct points are afforded distinct labels. There are a variety of parameters with these features that one could choose, but one standard choice is the \emph{generalized affine parameter}, to be discussed below. If we accept the generalized affine parameter as the right choice for the job, we can say that two space-like surfaces are finitely separated one from another just in case all of the time-like and light-like curves between the two surfaces have finite generalized affine length. We can then use this conception of a finite spacetime region in order to develop the notion that the Cosmos has a finite initial segment.

The fact that light-like trajectories record zero proper time can be derived from the fact that relativistic spacetimes are described by a spacetime interval with Lorentzian signature. Due to the Lorentzian signature, there are trajectories on which the temporal and spatial terms in the spacetime interval exactly cancel. In contrast, Euclidean spaces are described by positive definite metrics, so that there are no trajectories on which the spatial and temporal terms in the corresponding interval cancel. Our problem would be resolved if we could construct an appropriate map $\phi$ from any spacetime $S$ with Lorentzian signature to a space $S'$ with Euclidean signature with the following features: 

\begin{enumerate}
    \item For any finitely long time-like curve $I$ in $S$, the image of $I$, i.e., $\phi(I)$, in $S'$ is finitely long.
    \item For any infinitely long time-like curve $I_{\infty}$ in $S$, $\phi(I_{\infty})$ is infinitely long.
    \item For any two numerically distinct points, $p_1$ and $p_2$, connected by light-like curve $I_{\ell}$ in $S$, $\phi(I_{\ell})$ is non-zero in length.
    \item For any two numerically identical points, $p_1$ and $p_2$, in $S$, the respective images of $p_1$ and $p_2$ in $S'$, i.e., $\phi(p_1)$ and $\phi(p_2)$, are numerically identical and are separated by zero distance in $S'$.
\end{enumerate}

\noindent The generalized affine parameter, as defined below, is a natural choice that satisfies all four desiderata.

A \emph{half-curve} is usually defined as a curve that starts somewhere in spacetime and is inextendable. A classical spacetime model $S$ is said to be extendable just in case there is another larger spacetime model $S'$ into which $S$ can be isometrically embedded; moreover, $S$ is inextendable just in case $S$ is not extendable. A typical assumption in relativistic physics is that spacetimes are ``as large as they can be''; that is, that spacetime is inextendable. A curve $\gamma$ in $S$ is inextendable just in case there is no larger spacetime $S'$ into which $S$ can be isometrically embedded and in which $\gamma$ is longer than $\gamma$ was in $S$. Intuitively, an inextendable curve is a curve that encounters an impassible boundary to spacetime. For the sake of complete generality in explicating the concept of the beginning of the Cosmos, I will not assume that the Cosmos is inexteendable and, for that reason, I will offer a modified definition of half-curves. For my purposes, a half-curve in a spacetime $S$ is a curve that begins somewhere in $S$ and that has no further extension in $S$. Intuitively, if a half-curve $\gamma$ in $S$ has finite length, then $\gamma$ encounters a boundary of $S$.

Consider a classical spacetime represented by $(M, g)$. Without loss of generality, and utilizing the notation from \cite[p. 35]{Earman:1995}, consider a time-like half curve $\gamma(v)$ defined on $[0, v_+) \rightarrow M$, where $v$ is a parametrization of $\gamma$ and such that $v_+ \leq +\infty$. For each of the tangent spaces at each point in $M$, we can choose a set of four orthonormal basis vectors; the assignment of four orthonormal basis vectors at each of the spacetime points comprises the so-called ``frame field''. In particular, let's denote the basis vectors defined at each of the tangent spaces at each point along $\gamma(v)$ as $e^a_i(v)$, so that at $v=0$, the basis vectors are given by $e^a_i(0)$. Given $e^a_i(0)$, we can define the other basis vectors in the tangent spaces at the other points along $\gamma(v)$ via parallel transport.

Now that we have defined orthonormal basis vectors for each of the tangent spaces along $\gamma(v)$, we can write the components of a tangent vector $\mathbf{V}$ in terms of the $e^a_i(v)$ as:

$$
    V^a = \sum_{i=1}^4 X^i(v) e_i^a(v)
$$

\noindent The Euclidean length of $V^a$ is given by:

$$
    \vert \textbf{V} \vert = \sqrt{\sum_{i=1}^4 (X^i(v))^2}
$$

\noindent And, thus, we have succeeded in expressing the tangent vectors along $\gamma(v)$ using the Euclidean signature. Given the components of this tangent vector, we can write the generalized affine parameter $\lambda(v)$ as

$$
    \lambda(v) = \int_0^v \sqrt{\sum_{i=1}^4 (X^i(v^*))^2} dv^*
$$  

\noindent Where $v^*$ is a dummy variable replacing $v$ inside the integral. Since the summation under the square root within this integral is defined using a positive definite signature, the generalized affine parameter can be thought of as the arc length of a curve in a four-dimensional space instead of a four-dimensional spacetime. Using the generalized affine parameter, we can define a notion of \emph{generalized affine length}. The generalized affine length g.a.l. is the total length of $\gamma(v)$, that is,

$$
    \textnormal{g.a.l.} = \int_0^{v_+} \sqrt{\sum_{i=1}^4 (X^i(v^*))^2} dv^*
$$

\noindent As Earman notes, the choice of a different set of basis vectors $e^a_i(v)$ for each tangent space leads to a different generalized affine parameter defined on $\gamma(v)$. (Of course, once a choice of basis has been made at $v=0$, that choice can be propagated to every other point along $\gamma(v)$ by parallel transport.) But if one choice of basis vectors leads to a finite generalized affine length, then any other choice of basis vectors will lead to a finite generalized affine length; likewise, if any choice of basis vectors leads to infinite generalized affine length, then any other choice will lead to infinite generalized affine length. For that reason, whether the generalized affine length is finite or infinite is independent of our choice of orthonormal basis vectors and satisfies the desiderata identified at the end of the previous section.

Note that two space-like surfaces can be said to be finitely separated from each other just in case all of the time-like and light-like curves between the two surfaces have finite generalized affine length; this observation helps to illuminate the notion of a space-time region with finite size. Likewise, for the sake of intuition, imagine a closed or open boundary $B$, where the points in $B$ are all to the causal past of spacetime points not included in $B$, and a space-like surface $\Sigma$, all of whose points are in the causal future of points included in $B$. Suppose, further, that the Cosmos satisfies the Direction Condition alongside whatever other conditions are necessary for the Cosmos to have had a beginning. If all of the time-like and light-like curves between $B$ and $\Sigma$ have finite generalized affine length, then, intuitively, $B$ should count as the spacetime's beginning. This intuition turns out not to be precisely correct, but, as a first pass, we may say:

\bigskip

\noindent \emph{The Cosmos has a finite initial segment just in case there is a Cosmos-wide space-like surface $\Sigma$ such that all of the time-like and light-like curves that can be traced backwards from $\Sigma$ have finite generalized affine length.}

\bigskip

\noindent However, there is at least one way in which this conception of a finite initial segment of the Cosmos is inadequate. Recall that, in discussing a topological beginning, I discussed the possibility that there could be two or more metrics that objectively (or fundamentally) described spatio-temporal distances.\footnote{What do I mean by the `objective' or `fundamental' qualifier? Recall that one way to motivate a bimetric theory takes inspiration from Poincar{\'e}'s thought experiments in which there is an apparent metric due to the presence of universal forces on our measuring devices. By an `objective' or `fundamental' metric I mean a metric that is not merely apparent and that truly has metaphysical significance. I do not adopt a stance in this article on what precisely is requird for a metric to truly have metaphysical significance.} In that case, an initial segment of the Cosmos may be indeterminately old, in the sense that the initial segment might be finitely old with respect to one metric and infinitely old with respect to another. Moreover, since the generalized affine length is calculated with respect to a metric, on theories with two or more metrics, one may specify two or more generalized affine lengths. Consequently, we should add to the notion that the Cosmos has a finite initial segment the condition that the generalized affine length, when calculated with respect to all of the objective/fundamental spacetime metrics, yields a finite result:

\bigskip

\noindent \emph{The Cosmos has a finite initial segment just in case (i) there is one or more objective spacetime metrics and (ii) a Cosmos-wide space-like surface $\Sigma$ such that all of the time-like and light-like trajectories that can be traced backwards from $\Sigma$ have finite generalized affine length when calculated with respect to all of the objective spacetime metrics.}

\bigskip

\noindent There are two more worries that need to be overcome in order to state a completely satisfactory conception of a finite initial segment. Recall that in the Partially Amorphous Cosmos, the initial portion of spacetime might be described by an objective metric, even though a subsequent portion cannot be described by an objective metric. In that case, the Cosmos would have ``one or more objective spacetime metrics'' but only in the initial segment. Second, the definition I've offered for a finite initial segment assumes that the finite initial segment is bounded, in the future direction, by a space-like surface $\Sigma$. But if the future boundary of the initial segment is ``jagged'', there might be no space-like surface that bounds the initial segment. Let's say that a set $N$ of points is \emph{spacetime-wide} just in case every spacetime point not included in $N$ is in the causal future or the causal past of points included in $N$. Thus:

\bigskip

\noindent \emph{The Cosmos has a finite initial segment just in case: (i) There is a spacetime-wide set $\Sigma$; (ii) There is a region $R$ to the past of $\Sigma$; (iii) There are no points to the past of $\Sigma$ not contained in $R$; (iv) In $R$, there is at least one objective spacetime metric relating any two points in $R$; (v) All of the time-like and light-like curves that can be traced backwards from $\Sigma$ have finite generalized affine length when calculated with respect to all of the objective spacetime metrics in $R$.}

\bigskip

\noindent In turn, the Metrical Conception states that the Cosmos has a metrical beginning only if the Cosmos has a finite initial segment. And then the Boundary Condition can be completed by taking the disjunction of the Topological Conception and the Metrical Conception, i.e., in order for the Cosmos to have a beginning, the Cosmos must satisfy the Topological Conception or the Metrical Conception.

One may worry that I have made three implicit assumptions in explicating what it would mean for the initial segment of the Cosmos's history to be finite. First, one might worry that I have assumed spacetime to be a continuum. To the contrary, if spacetime is discrete, then one of two conditions will be satisfied. Either the past of every non-initial spacetime atom includes a first spacetime atom (or atoms) so that the Cosmos satisfies the Topological Conception or else the Cosmos simply does not have a beginning.\footnote{The first spacetime atom in each temporal series of spacetime atoms need not be numerically identical to the first spacetime atom in any other temporal series of spacetime atoms. Note that this allows for an interesting possibility. There could be a discrete spacetime with a jagged edge, i.e., a discrete spacetime such that there are only a finite number of spacetime atoms to the past of every non-initial spacetime atom and yet there is no upper bound to the number of spacetime atoms to the past of non-initial spacetime atoms. For example, for every spacetime atom with $n$ previous atoms, there may be another with $n+1$ previous atoms.} Thus, we do not need a Metrical Conception of the Cosmos's beginning. But, even supposing that we do need a Metrical Conception of the Cosmos's beginning in the case where spacetime has a discrete structure, the generalized affine length has a natural extension in terms of the counting measure. 

Second, one might worry that I've assumed -- without argumentation -- that the generalized affine lengths of curves are incommensurate or otherwise not objectively comparable. However, this worry can be assuaged in the following way. The Metrical Conception requires \emph{only} that all of the generalized affine lengths of curves traced back backwards from $\Sigma$ are finite and not that the specific numerical values assumed by those lengths are directly comparable.

Third, one might worry that I have assumed that there is a meaningful distinction between finite and infinite spacetime regions. For example, if the lengths of temporal durations are conventional or if time is amorphous in the initial segment, then there is no objective distinction between finite and infinite initial segments. But this worry is mistaken. On the one hand, if there is no objective finite/infinite distinction in the initial segment and the initial segment has a closed boundary, then (so long as the Cosmos satisfies whatever other conditions there may be for the Cosmos's beginning) the initial segment has a topological beginning. Since the Boundary Condition is disjunctive, nothing, at least in terms of the boundary, would block us from saying that the Cosmos has a beginning. On the other hand, if there is no objective finite/infinite distinction in the initial segment and the initial segment has an open boundary, then the initial segment is not finite. In that case, the Cosmos would not satisfy the Boundary Condition and would not have a beginning.

\section{\label{objections}Objections}

In this section, I turn to two important objections to the view that I've presented in this article. According to the first objection, while the Boundary Condition, as I've stated it, captures two of the ways in which spacetime could have a boundary, I haven't shown that there are no other ways in which spacetime could have a boundary. According to the second objection, one or the other of the two disjuncts -- i.e., either the Topological Conception or the Metrical Conception -- is broader and should subsume the other. That is, the second objection challenges us to say why we need both disjuncts and cannot make do with only one. In the following, I answer both objections and show that both objections fail.

\subsection{The First Objection: Uniqueness?}

On my view, the Boundary Condition is a necessary condition for the Cosmos to have a beginning. While the reader might share my intuition that the Cosmos having a beginning requires that the Cosmos include a past boundary of some kind, the Boundary Condition -- at least as I stated the condition -- can be a necessary condition for the Cosmos to have a beginning only if the Topological Conception and the Metrical Conception exhaust all of the relevant ways for the Cosmos to have a past boundary. Why think that the Topological Conception and the Metrical Conception are the only two relevant ways for the Cosmos to include a past boundary? According to a standard mathematical procedure for constructing a space or spacetime, we begin with a set of simples (or ``points'') which we can then endow with additional structure, e.g., \cite{Norton:1999}; \cite[p. 10--11]{Isham:1993}; \cite{Maudlin:2010}, \cite[p. 5-8]{Maudlin:2012}; \cite[p. 14-18]{DeLanda:2013}; \cite[p. 40-51]{North:2021}. The additional structure forms a hierarchy, that is,

\begin{enumerate}
    \item\hspace*{1sp} The \emph{set theoretic structure} describes the properties the point set has in virtue of being a set, e.g., the cardinality of the point set or whether a given entity is a member of the point set. 
    \item\hspace*{1sp} The \emph{topological structure} describes the continuity or discontinuity of the space or spacetime as well as whether the space or spacetime has closed, open, or partially open boundaries.
    \item\hspace*{1sp} The \emph{affine structure} describes the primitive distinction between curves and straight lines.
    \item\hspace*{1sp} The \emph{metrical structure} describes the distance (or interval) between any two points.
    \item\hspace*{1sp} The \emph{differentiable structure} allows us to distinguish smooth curves from curves with sharp or broken edges.
\end{enumerate}

\noindent Additional structure can be defined on any given point set as well. For example, A-theories of time define primitive temporal structure in terms of the monadic predicates of pastness, presentness, and futurity. Consequently, A-theories endow spacetime with what I will call \emph{monadic structure}. On B-theories of time, a binary relation -- the B-relation -- is defined between any two numerically distinct time-like related events $\alpha$ and $\beta$, in virtue of which we can say either that $\alpha$ is before $\beta$ or $\beta$ is before $\alpha$. Likewise, on some -- albeit outdated -- metaphysical accounts of the nature of space (or of the nature of place), we should supplement space with additional structure. For example, Aristotle's view of the nature of place denies the homogeneity of space and defines the center of the Earth as the center of the Cosmos. For that reason, Aristotle's view includes fundamental and irreducible relations of \emph{up} and \emph{down}. Let's call the additional structure added in the case of either B-theory or the Aristotelian conception of place \emph{ordinal structure}, since, in either case, we are imposing an ordering relation on a given point set.

Plausibly, the Boundary Condition should be definable in terms of the formal structure out of which we can construct models of spacetime. Intuitively, given the various formal structures described above, only two kinds of formal structure -- that is, topological structure and metrical structure -- are capable of capturing the notion of a boundary. For example, when we say that an ordinary object, e.g., a table, has a boundary, we might mean that, e.g., the table has an edge, that is, a topological boundary, or we might mean that the table has finite spatial extension, that is, a metrical boundary. We don't mean that the table has a boundary in virtue of our ability to define straight lines on the table, or our ability to distinguish smooth curves from curves with sharp edges, or in terms of some ordinal or monadic structure that we can define on the parts of the table.\footnote{Perhaps the reader will object that one way that a series can have a boundary involves the series having a first member and having a first member has to do with the ordinal structure of the series. Nonetheless, the conjunction of the Direction Condition and the Boundary Condition successfully captures the Cosmos's ordinal structure; for example, the Cosmos might have a closed boundary -- and so satisfy the Topological Conception -- and that closed boundary might be the Cosmos's first moment in virtue of satisfying the Direction Condition. Moreover, the conjunction of the Direction Condition and the Boundary Condition capture a broader range of cases than if we defined the Boundary Condition in terms of the ordinal structure of spacetime.} Since there are only two ways of capturing the notion of a boundary in terms of the formal structure out of which we can construct models of spacetime, I've defined the Boundary Condition disjunctively in terms of those two notions.

\subsection{The Second Objection: Disjunctive or Atomic?}

According to the second objection, one or the other of the two disjuncts -- i.e., either the Topological Conception or the Metrical Conception -- is broader and should subsume the other. If so, the Boundary Condition would be atomic instead of disjunctive. In order to dissuade readers from this objection, I will present an example of a family of spacetimes that satisfy the Topological Conception but not the Metrical Conception as well as an example of a family of spacetimes that satisfy the Metrical Conception without satisfying the Topological Conception. In light of the two examples, I will conclude that we need both the Topological Conception and the Metrical Conception.

Let's begin with an example of a family of spacetimes satisfying the Topological Conception but not the Metrical Conception. Here, we can consider any spacetime with an initial closed boundary that has an initial segment that is metrically amorphous. Since the spacetime has an initial closed boundary, the spacetime satisfies the Topological Conception. But since the spacetime is initially metrically amorphous, the spacetime lacks a finite initial segment. Now, let's turn to an example of a family of spacetimes satisfying the Metrical Conception but not the Topological Conception. Any spacetime with a finite initial segment but an open boundary will satisfy the Metrical Conception but fail to satisfy the Topological Conception. For example, singular FLRW spacetimes satisfy the Metrical Conception but not the Topological Conception.

\begin{comment}
    There is a second way that this objection can be pressed. Perhaps some readers will not be persuaded that one of the disjuncts helps to capture a notion of the Cosmos's beginning; if so, they may say that we ought to define the Boundary Condition solely in terms of one of the disjuncts but not the other. In reply, recall that I am looking for a set of necessary conditions for the Cosmos to have had a beginning. If one or the other of the two disjuncts in the Boundary Condition captures a necessary condition for a beginning while the other does not, then the disjunction of the two will still be a necessary condition. Given that a definition in terms of the disjunction is logically weaker than a definition in terms of either disjunct, a definition in terms of the disjunction of the two is more likely to be correct than a definition in terms of either taken individually.
\end{comment}

There is a second way to state the objection considered in this subjection. Some readers may not agree that one of the two options I presented successfully captures a notion of the Cosmos's beginning. They might suggest that we should only use one, but not the other, of the disjuncts to define the Boundary Condition, in which case the Boundary Condition should be defined atomically and not disjunctively. However, I am looking for a set of necessary conditions for the beginning of the Cosmos. If one of the options is necessary while the other is not, then the disjunction of the two will still be necessary. In virtue of being logically weaker, a disjunction is more likely to be correct than a definition in terms of only one of the two disjuncts.

\section{\label{big-bang}The Cosmic Boundary and Classical Big Bang Models}

Various authors have expressed the intuitive idea that if classical Big Bang models were correct, then the Cosmos would have a beginning. There may be space for disagreement on this point. For example, at the outset of this paper, I mentioned the possibility that another condition -- such as the truth of a tensed theory of time -- may be necessary for the Cosmos to have had a beginning. However, many of the authors who have written on classical Big Bang models appear to have assumed that no conditions other than that the entire Cosmos has a boundary to our past -- that is, the Direction and Boundary Conditions -- are required for the Cosmos to have had a beginning. Consequently, we should interpret the claim that classical Big Bang models involve a beginning as the claim that classical Big Bang models satisfy the intuitive notions that the the Direction and Boundary Conditions are meant to capture. Moreover, insofar as there is a commonplace claim that classical Big Bang models satisfy the intuitive notions that the the Direction and Boundary Conditions are meant to capture, we can test my articulation of those two conditions against classical Big Bang models. If we cannot come to an intuitive understanding of the relationship between classical Big Bang models and the Direction/Boundary Conditions as I've articulated them, then we would have good reason to think that the Direction/Boundary Conditions do not successfully capture necessary conditions for the Cosmos to have had a beginning.

In order to support my articulation of the Direction/Boundary Conditions, I prove two closely related results in this section. First, I will prove that all maximally extended classical spacetimes that satisfy the Direction and Boundary Conditions are b-incomplete. Second, if (i) General Relativity is true, (ii) the cosmological principle is true, (iii) spacetime is maximally extended, (iv) spacetime satisfies the Direction Condition, and (v) spacetime satisfies the Boundary Condition, then spacetime is correctly modeled by one of the FLRW metrics with a Big Bang singularity. In other words, the Direction and Boundary Conditions, together with some other assumptions, allow one to \emph{derive} classical, singular Big Bang models. The two short theorems precisify and modify the intuitive notion that classical, singular Big Bang models include a beginning and, in doing so, provide evidence that I have correctly provided two necessary conditions for the Cosmos to have a beginning. The two theorems also support the view that the Direction and Boundary Conditions are more general (and so more fundamental) than the sort of ``beginning'' involved in classical Big Bang models.

\subsection{B-Incompleteness and Singular Spacetimes}

The notion that classical Big Bang models, if interpreted literally, entail a beginning of the Cosmos is closely related to the notion of a spacetime singularity. Unfortunately, physicists, philosophers of physics, and mathematicians have yet to develop a fully satisfactory set of conditions for distinguishing singular from non-singular spacetimes. Given the deeply technical nature of this problem, the solution is beyond my current abilities and I will not attempt to resolve the problem here. Instead, I will summarize some of the relevant literature in order to offer one standard, if not fully satisfactory, conception of how singular and non-singular spacetimes differ.

Although singular FLRW spacetimes include a divergent Ricci scalar, divergences in the various curvature parameters are neither necessary nor sufficient for a classical spacetime to be singular \cite{Earman:1995, Curiel:1999, sep-spacetime-singularities, Joshi:2014}. For my purposes, we can utilize what John Earman \cite[p. 36]{Earman:1995} calls the ``semi-official definition'' and what elsewhere has been called the ``most widely accepted solution'' for defining singular spacetimes \cite{sep-spacetime-singularities}. A classical spacetime is said to be \emph{b-complete} just in case every time-like and light-like half-curve has infinite generalized affine length. According to Earman's semi-official definition, a classical spacetime is then said to be singular just in case the spacetime is not b-complete. Arguably, one should add the condition that spacetime is maximally extended \cite[p. 715]{Lam:2007}. Since this definition is not completely satisfactory,\footnote{For some of the problems involved with utilizing b-incompleteness as the definitive feature of singular spacetimes, see chapter $2$ in \cite{Earman:1995}; also see \cite{sep-spacetime-singularities}.} I will not take up the position here that all and only singular spacetimes are b-incomplete. Moreover, I will not take up the debate, e.g., \cite[p. 32]{Earman:1995}; \cite{Manchak:2021}, as to whether the spacetime we inhabit is maximally extended. Instead, I will assume that spacetime is maximally extended. In any case, b-incompleteness will allow us to see the precise sense in which the Direction and Boundary Conditions relate to singular spacetimes. That is, as I prove in the next section, all classical spacetimes that are maximally extended and that satisfy the Direction and Boundary Conditions are b-incomplete.

\subsection{The Two Theorems}

\textbf{Theorem 1.} All maximally extended classical spacetimes that satisfy the Direction and Boundary Conditions are b-incomplete.

\textbf{Proof.} To begin the proof, let's assume that the Cosmos includes a maximally extended classical spacetime satisfying the Direction and Boundary Conditions. Recall that, according to the Boundary Condition, the Cosmos began to exist just in case either there is a Cosmos-wide closed boundary to the past of every non-initial spacetime point or there is an initial objectively finite portion of the Cosmos’s history. We can proceed to prove by cases. 

Let's first suppose that spacetime has a closed boundary $\mathcal{B}$ to the past of every non-initial spacetime point.\footnote{Here, $\mathcal{B}$ need not be a space-like surface. Instead, I will understand $\mathcal{B}$ as the set of closed initial points for all of the time-like and light-like curves in the spacetime. In the case of a spacetime with a ``jagged'' boundary, there may be no simple relationship -- and possibly no continuity -- between the points in the set.} The proof for this case is trivial. Consider any time-like or light-like half-curve $\gamma$ that originates at some point $p \in \mathcal{B}$. Any such curve will have zero extension backwards through the spacetime.\footnote{This result will not necessarily follow for any curve that is not located in $\mathcal{B}$. For example, suppose that spacetime has a closed boundary but that the initial portion of the Cosmos has the ``fractal'' metrical properties discussed above. In that case, any time-like or light-like curve not located in $\mathcal{B}$ will have infinite backwards extension.} Since the curve has zero backwards extension, the spacetime is b-incomplete. Having established the first case, let's move to the second. Suppose that there is an initial objectively finite portion of the Cosmos's history. Now consider an arbitrary time-like or light-like half-curve intersecting some spacetime-wide set $\Sigma$ in the initial objectively finite portion and that extends backwards through the Cosmos. By the definition of an objectively finite portion of the Cosmos's history established above, this half-curve must have finite generalized affine length. 

Therefore, if the Cosmos includes a maximally extended classical spacetime satisfying either of the two disjuncts -- and so satisfying the Boundary Condition -- the Cosmos is b-incomplete. Consequently, we have the desired result, i.e., if the Cosmos includes a maximally extended classical spacetime satisfying the Direction and Boundary Conditions, then spacetime is b-incomplete. 

\textbf{Discussion.} Two cautionary notes are in order. First, the converse of the first theorem does not hold, i.e., if the Cosmos is b-incomplete, it would not follow that the Cosmos satisfies the Direction and Boundary Conditions. By this point in this article, the reason should be obvious. If the Cosmos were b-incomplete, this would tell us, at most, that at least one time-like or light-like half-curve has finite generalized affine length. \emph{One} way that at least one time-like or light-like half-curve could have finite generalized affine length involves the Cosmos satisfying the Boundary Condition. In that case, the Cosmos might still fail to satisfy the Direction Condition or perhaps other conditions necessary for the Cosmos to have had a beginning. Of course, the Cosmos could fail to satisfy the Boundary Condition even if at least one time-like or light-like half-curve had finite generalized affine length, since, in that case, there may be other time-like or light-like half curves that do not have finite generalized affine length. Moreover, even if spacetime were finite to the past, with no extension to the past of the Big Bang, the Cosmos might still fail to satisfy the Direction Condition (or perhaps other conditions necessary for the Cosmos to have had a beginning) and so fail to have a beginning.

Arguably, the Direction and Boundary Conditions are more fundamental than the Big Bang singularity because the Direction and Boundary Conditions figure into a derivation of the Big Bang singularity. In order to see that this is so, I turn to the second theorem. 

\textbf{Theorem 2.} If (i) General Relativity is true, (ii) the cosmological principle is true, (iii) spacetime is maximally extended, (iv) spacetime satisfies the Direction Condition, and (v) spacetime satisfies the Boundary Condition then spacetime is correctly modeled by one of the FLRW metrics with a Big Bang singularity.

\textbf{Proof.} Suppose (i)-(v) are true. Assumption (i) -- that General Relativity is true -- entails the Einstein Field Equations. The Einstein Field Equations together with assumption (ii), i.e., that the cosmological principle is true -- that is, that spacetime can be ``cut up'' (or foliated) into space-like surfaces on which the matter-energy distribution is homogeneous and isotropic -- entails that spacetime is one of the FLRW models. Since we've assumed that spacetime is maximally extended (assumption (iii)), we can make the further restriction to maximally extended FLRW models. Maximally extended FLRW models can be subdivided into two families: those that include a Big Bang type singularity and those that do not. Using the fact that the Direction and Boundary Conditions together entail that spacetime is b-incomplete (i.e., theorem 1), we can eliminate the FLRW models that do not include a Big Bang type singularity. Thus, in the context of General Relativity, the Direction and Boundary Conditions, together with some additional assumptions about the global structure of spacetime and the matter-energy distribution, can be used to derive the Big Bang singularity.

\textbf{Discussion.} This result—that the Direction and Boundary Conditions can figure into a derivation of the Big Bang singularity—helps to show one sense in which the Direction and Boundary Conditions are more fundamental than Big Bang theory. The result also helps to clarify why, on the assumption that General Relativity is true, one still cannot infer that the Cosmos began to exist. The derivation of the Big Bang singularity from the Direction and Boundary Conditions utilized an unre- stricted version of the cosmological principle. One may be able to make use of some other global property. However, Clark Glymour \cite{Glymour:1972, Glymour:1977}, David Malament \cite{Malament:1977}, J.B. Manchak \cite{Manchak:2009, Manchak:2011, Manchak:2021, Manchak:2021b},  Claus Beisbart \cite{Beisbart:2009, Beisbart:2022}, John Norton \cite{Norton:2011}, and Jeremy Butterfield \cite{Butterfield:2014} have previously challenged the notion that any observer, embedded within a relativistic spacetime, could gather enough data to infer that their spacetime satisfies various global properties, including an unrestricted version of the cosmological principle. Additionally, several live cosmological models (e.g., models featuring an inflationary multiverse) entail that the matter-energy distribution is not homogeneous and isotropic on scales significantly larger than our Hubble volume. In order to know whether the Direction and Boundary Conditions are satisfied for the entire Cosmos, one would need to know substantive details about the global distribution of matter-energy that no one could be in an epistemic position to know.

\section{Conclusion}

I will conclude by summarizing the road we've followed. We began with the intuitive notion that an entity begins to exist just in case there is a temporal boundary before which the entity did not exist. This intuition needs to be made more precise. While Pitts \cite{Pitts:2008} previously offered a useful distinction between the topological and metrical senses of a beginning, I have shown that his version of the Metrical Conception is inadequate. The novel proposal that I offered in this paper borrows Pitts's distinction, improves on the Metrical Conception, and is defined in terms of a disjunction between the two. According to my proposal, the Cosmos had a beginning only if either the Topological Conception or the Metrical Conception are satisfied. According to the Topological Conception, there is a closed boundary to the past of every non-initial spacetime point. According to the Metrical Conception, there is an initial objectively finite portion of the Cosmos’s history. In turn, there is an initial finite portion of the Cosmos's history just in case: (i) There is a spacetime-wide set $\Sigma$; (ii) There is a region $R$ to the past of $\Sigma$; (iii) There are no points to the past of $\Sigma$ not contained in $R$; (iv) In $R$, there is at least one objective spacetime metric relating any two points in $R$; (v) All of the time-like and light-like curves that can be traced backwards from $\Sigma$ have finite generalized affine length when calculated with respect to all of the objective spacetime metrics in $R$.

Lastly, I demonstrated a relationship between classical, singular Big Bang models and the Direction and Boundary Conditions. I proved two theorems that explain the intuition that classical, singular Big Bang models involve a beginning. According to the first theorem, any classical and maximally extended spacetime satisfying the Direction and Boundary Conditions is b-incomplete. According to the second theorem, if (i) General Relativity is true, (ii) the cosmological principle is true, (iii) spacetime is maximally extended, (iv) spacetime satisfies the Direction Condition, and (v) spacetime satisfies the Boundary Condition then spacetime is correctly modeled by one of the FLRW metrics with a Big Bang singularity. Due to the way in which the Direction and Boundary Conditions figure into a derivation of the Big Bang singularity, I concluded that the Direction and Boundary Conditions are more fundamental to the notion that the Cosmos has a beginning than the Big Bang singularity. Moreover, the derivation helps to explain the common intuition that the Big Bang singularity ought to be interpreted as the universe's beginning.

%\bibliography{references.bib}
\bibliographystyle{plainnat}
\bibliography{references.bib}

\end{document}